\documentclass[a4paper,10pt]{amsart}
\usepackage{xcolor}
\usepackage[utf8]{inputenc}
\usepackage[a4paper, left=4.0cm, right=4.0cm, top=3.0cm, bottom=3.5cm, includefoot]{geometry}
\usepackage[
    hyperref, backend=biber, doi=false,url=false,sorting=none,backref=true]{biblatex}
\DefineBibliographyStrings{spanish}{andothers={\textsl{et al}\adddot}}
\renewbibmacro{in:}{}

\usepackage{amsmath, amssymb,stackrel,amsaddr}
\usepackage{hyperref}
\usepackage[english]{babel}
\usepackage{graphicx}

\hypersetup{colorlinks=true,linktoc=page,linkcolor=blue,citecolor=red,urlcolor=cyan}

\bibliography{bibliografia}

\title{Asymptotic freedom for $\lambda \phi^4_{\star}$ QFT in Snyder-de Sitter space}

\author{S.~A.~Franchino-Vi\~nas}
\address{Departamento de F\'isica, Facultad de Ciencias Exactas
Universidad Nacional de La Plata, C.C.\ 67 (1900), La Plata, Argentina,}
\address{Theoretisch-Physikalisches Institut, Friedrich Schiller Universit\"at Jena, Max Wien Platz 1, 07743 Jena, Germany.}
\email{sa.franchino@uni-jena.de}
\author{S. Mignemi}
\address{Dipartimento di Matematica e Informatica, Università di Cagliari, viale Merello 92, \\09123 Cagliari, Italy,}
\address{INFN, Sezione di Cagliari, Cittadella Universitaria, 09042 Monserrato, Italy.}
\email{smignemi@unica.it}

\begin{document}

\begin{abstract}
We analyze the model of a self-interacting $\phi^4_{\star}$ scalar field theory in Snyder-de Sitter space.
After analytically computing the one-loop beta functions {in the small noncommutativity and curvature limit},
we solve numerically the corresponding system of differential equations,
showing that in this limit the model possesses at least one regime in which the theory is asymptotically free.
Moreover, in a given region of the parameter space we also observe a peculiar running of the parameter associated to the curvature,
which changes its sign and therefore can be interpreted as a transition from an IR de-Sitter space to and UV anti-de Sitter one.
\end{abstract}
\maketitle

\section{Introduction}
It is currently believed that noncommutative geometry \cite{Connes:1994,Doplicher:1994tu} may play an important role in the search for a quantum theory of gravity.
The first example of noncommutative geometry was introduced by Snyder \cite{Snyder:1947nq} with the hope of taming the divergences of quantum field theory (QFT) through the discretization of space. Of particular interest is therefore the investigation of QFT on noncommutative spaces, which has become an important area of research in recent times, especially in the context of the Moyal geometry \cite{Douglas:2001ba,Szabo:2001kg}.

In \cite{Franchino-Vinas:2018jcs}, we have used the worldline formalism to investigate the QFT of a scalar model with a quartic interaction on Snyder space,
in an approximation of first order in the noncommutativity parameter.
The investigation has been performed using the standard formalism of noncommutative QFT.
{We have found that divergences appear in the six-point function which could lead to renormalization problems.
However, this is due to the approximation used.} A treatment at all orders in the noncommutativity parameter
gives strong clues of  renormalizability and maybe even finiteness,
at least for some choices of the interaction potential,
although calculations have not been completed due to algebraic difficulties \cite{Meljanac:2017jyk}.

%However, the present status of these calculations cannot rule out the occurrence of the phenomenon of UV/IR mixing.
{An intriguing phenomenon that occurs in several examples of noncommutative QFTs is that of UV/IR mixing \cite{Minwalla:1999px}.}
It manifests itself through the appearance of divergences in the UV-renormalized diagrams when
the external momenta go to zero. Although this scenario could compromise the renormalizability of the theory, it also brings out several opportunities.
The mixing could be for example used to generate small scales from the UV dynamics as pointed out
in \cite{Craig:2019zbn}. It also served as a motivation to find new well-behaved NC field theories.
{One of the major successes in this respect} is the Grosse-Wulkenhaar (GW) extension of the Moyal $\phi^4_\star$ model \cite{Grosse:2012uv}.
The main peculiarity of this model is the presence in the action of an harmonic oscillator term that
smooths the infrared behavior and in the end gives rise to an all-order perturbatively renormalizable theory.
{According to the latest results, in a certain limit it could provide the first example
of a solvable model in four dimensions \cite{Grosse:2019qps}.}

It is therefore naturally of interest to further investigate the origin of the harmonic term. It has been observed that the introduction of such term may be justified
by the assumption of a de Sitter spacetime  {background \cite{Franchino-Vinas:2019lyi}}.
For this reason, we extend our previous study of the Snyder scalar QFT to the case of a de Sitter background. Some other approaches to noncommutative curved spaces can be found in \cite{Mignemi:2008fj,Gutierrez-Sagredo:2019ipf,Ballesteros:2019hbw}, where the authors motivate their mathematical construction from Poisson-Lie algebras, and \cite{Buric:2019yau,Buric:2018kdi}, where a group-representation approach is used and some astrophysical consequences are discussed.

Coming back to our Snyder-de Sitter model, we shall show that it presents an effective action that at zeroth order in the
noncommutative and curvature parameters coincides with that of the GW model (although the star-product is different), and might then share some of its properties.
{Another property common to the GW and Snyder-de Sitter algebra is the existence of a duality between positions and momenta. However, as we will see it is difficult to extend the duality to the level of the action in the latter case.}

{Our computations are limited to the first approximation in the noncommutativity and curvature coupling constants of the theory and therefore are just a starting point, which permits to study the {renormalization group (RG)} flow of the theory. We will prove that at the one-loop level our model possesses several interesting properties: it admits asymptotic freedom in some region of the parameter space, as well as {a RG-induced change of sign }in the spacetime curvature.}

 Of course, the fact that we are working in a small-non\-com\-mu\-ta\-tivity expansion forbids us to drive conclusions about questions such as the UV/IR mixing. Also, for simplicity, we shall use an action which is not de Sitter symmetric. We shall present a fully de Sitter-invariant treatment together with a more detailed exposition of the calculations in a forthcoming paper. %\cite{franchinomignemi}
{For the same reason the question of (all-order) renormalizability is not tackled in this work and will be addressed elsewhere.
The model investigated in this paper is to be considered as an effective field theory whose features
could give us clues on what to expect from a noncommutative QFTs in a curved background, which at least in principle is more realistic than those usually considered in the literature.}

Coming back to our Snyder-de Sitter model, we shall show that it presents an effective action that at zeroth order in the
noncommutative and curvature parameters coincides with that of the GW model (although the star-product is different), and might then share some of its properties.
{Our computations are limited to the first approximation in the noncommutativity and curvature coupling constants of the theory and therefore are just a starting point, which permits to study the {renormalization group (RG)} flow of the theory. We will prove that at the one-loop level our model possesses several interesting properties: it admits asymptotic freedom in some region of the parameter space, as well as {a RG-induced change of sign }in the spacetime curvature.}

 Of course, the fact that we are working in a small-noncommutativity expansion forbids us to drive conclusions about questions such as the UV/IR mixing. Also, for simplicity, we shall use an action which is not de Sitter symmetric. We shall present a fully de Sitter-invariant treatment together with a more detailed exposition of the calculations in a forthcoming paper \cite{franchinomignemi}.
{For the same reason the question of (all-order) renormalizability is not tackled in this work and will be addressed elsewhere.
The model investigated in this paper is to be considered as an effective field theory whose features
could give us clues on what to expect from a noncommutative QFTs in a curved background, which at least in principle is more realistic than those usually considered in the literature.}

\section{Curved Snyder Space}
Snyder spaces are spaces in which noncommutativity is implemented in such a way that the Lorentz algebra and its action on the position ($\hat{x}_i$) and momentum ($\hat{p}_j$) operators is undeformed; in our Euclidean version, we can write the relevant commutation relations in terms of the generators
$J_{ij}$ of the Lorentz algebra,
\begin{align}
 \begin{split}
[J_{ij}, J_{kl}]&=i (\delta_{ik}J_{jl}-\delta_{il}J_{jk}-\delta_{jk}J_{il}+\delta_{jl}J_{ik}),\\
 [J_{ij}, \hat{p}_{k}]&=i (\delta_{ik} \hat{p}_{j}-\delta_{kj} \hat{p}_{i}),\\
 [J_{ij}, \hat{x}_{k}]&=i (\delta_{ik} \hat{x}_{j}-\delta_{kj} \hat{x}_{i}).
 \end{split}
 \end{align}
These expressions do not fix the commutation relations among momentum and position operators, that constitute the (deformed) Heisenberg algebra. Indeed, there exist several representations for which the $\hat{p}_i$ commute among themselves, but the $\hat{x}_i$ do not \cite{Battisti:2010sr}. The commutation relations of the $\hat{p}_i$ and $\hat{x}_i$ are then fixed almost uniquely by requiring the validity of the Jacobi identities.

In this paper, we consider instead an algebra in which the momenta do not commute: this can be interpreted as implying the presence of spacetime curvature. Explicitly, we choose  the commutation relations\footnote{Note the change $\beta\rightarrow \beta^2$ with respect to some previous works on the topic, as \cite{Meljanac:2017ikx}.} \cite{KowalskiGlikman:2004kp,Mignemi:2009zz,Mignemi:2011wh,Mignemi:2015una}
\begin{align}\label{eq:commutation}
\begin{split}
 &[\hat{x}_i,\hat{x}_j]=i\beta^2 J_{ij},\qquad [\hat{p}_i,\hat{p}_j]=i\alpha^2 J_{ij},\\
 &[\hat{x}_i,\hat{p}_j]=i[\delta_{ij}+\alpha^2 \hat{x}_i\hat{x}_j+\beta^2\hat{p_j}\hat{p}_i+\alpha\beta (\hat{x}_j\hat{p}_i+\hat{p_i}\hat{x}_j)].
 \end{split}
\end{align}

The generators $J_{ij}$ can be realized in terms of the phase space variables as $J_{ij}=\frac{1}{2}(\hat x_i\hat p_j-\hat x_j\hat p_i+\hat p_j\hat x_i-\hat p_i\hat x_j)$.
On the one hand, the position commutators correspond to the simplest Snyder space realization. On the other hand, the momentum commutators are the ones one would find for a de Sitter space in the usual commutative geometry. For $\beta\to 0$ one recovers the de Sitter algebra, while for $\alpha\to 0$ one gets the Snyder one. As usual, the commutator that mixes positions and momenta is fixed by the request that the Jacobi identities are satisfied. These observations justify the name ``Snyder-de Sitter space'' (SdS) given to this geometry. {It is also important to observe that the algebra \eqref{eq:commutation} is invariant under the duality 
\begin{align}\label{eq:duality}
\alpha\hat x_i \leftrightarrow\beta\hat p_i. 
\end{align}
 }

One interesting property of the SdS model  is the fact that it can be obtained from the usual Snyder space by means of a nonunitary transformation. In fact, we can define the operators $\hat{x}_i$ and $\hat{p}_i$ in terms of the operators $X_i$ and $P_i$ that satisfy the Snyder algebra,
\begin{equation}\label{Sny}
 [X_i,X_j]=i\beta^2 J_{ij}, \quad [P_i,P_j]=0,\quad [X_i,P_j]=i(\delta_{ij}+\beta^2P_iP_j),
\end{equation}
employing the following definitions that involve  an arbitrary parameter $t$ \cite{Mignemi:2009zz,Mignemi:2015una}, { which will be set to zero for reasons that will be explained below}:
\begin{align}\label{eq:2snyder}
 \hat{x}_i=:X_i+t\, \frac{\beta}{\alpha} P_i, &\qquad \hat{p}_i=:(1-t)P_i -\frac{\alpha}{\beta} X_i.
\end{align}
{In the following, we use this map in order to perform our calculations. 

It must be noted that this method has several disadvantages. However, at present there is no more practicable way to perform an investigation of the model.
For example, this map does not preserve the duality between positions and momenta of \eqref{eq:commutation},
i.e. there is no duality $a \alpha X_i \leftrightarrow b \beta P_i$ with $a,b\in \mathbb{R}$ such that  \eqref{eq:duality} is satisfied. In any case, also the choice of an action invariant under the full algebra \eqref{eq:commutation}, and hence under duality, is problematic.

Another disadvantage  of this representation is that it is singular for $\beta \to 0$. The reason is that the transformation \eqref{eq:2snyder} can give rise to the Snyder-de Sitter algebra  only if the operators $X_i$ are noncommutative, and hence $\beta\neq0$, i.e.~ the model must be thought as a deformation of Snyder space.

Certainly, there exist different representations for the algebra but we are not aware of any representation that solves the above problems. One possibility could be to start from the de Sitter algebra and repeat the previous steps in a dual way, obtaining therefore an expression with factors $\frac{\beta}{\alpha}$. The arousal of these non-anali\-ti\-cities can be understood in terms of a dimensional analysis: if we are willing to use a linear transformation as in \eqref{eq:2snyder} without the introduction of additional dimensional parameters, then the only possibility one can have is one of the already mentioned. If instead one relaxes one of the assumptions, we have no proof that forbids the existence of transformations that map the Snyder-de Sitter algebra into a Snyder algebra. We can just say that without the addition of extra dimensional parameters, there is no linear transformation with the desired properties. Also related to this, there are many known cases in noncommutative theories where the algebra shows a pole in some noncommutative parameter, usually related to the fact that the associated group modifies its compactness properties, giving place to the so-called quantum phases \cite{Nair:2000ii,Bellucci:2001xp}.}

Moreover, it is well known that the Snyder operators can be written in terms of operators $x_i$ and $p_i$, obeying canonical commutation relations, as
\begin{align}\label{eq:2canonical}
 P_i=:p_i=-i\partial_i, \quad X_i=:x_i+\beta^2x_jp_jp_i=x_i-\beta^2x_j\partial_j\partial_i.
\end{align}
Although the $X_i$ operators so defined are non-hermitian, this problem can be overcome by symmetrizing  \cite{Meljanac:2017ikx},
\begin{align}\label{eq:symmetrization}
 X_i\rightarrow X_i=\hat x_i=x_i+\frac{\beta^2}{2} (x_jp_jp_i+p_ip_jx_j).
\end{align}
After this sequence of transformations,  the original momentum operator of SdS can finally be written as
\begin{align}\label{eq:phat}
 \hat{p}_i=p_i-\frac{\alpha}{\beta} x_i-\frac{\alpha \beta}{2} (x_jp_jp_i+p_ip_jx_j).
\end{align}

\section{Self-interacting scalar quantum field theory on SdS}\label{sec:QFT}
In the following, we shall define the free scalar field action in $D$ dimensions as
\begin{align}\label{eq.kinetic}
 S_K=\frac{1}{2}\int d^Dx \ \varphi\, (\hat p^2+m^2)\, \varphi,
\end{align}
where $\hat p^2$ is the kinetic operator, defined previously in \eqref{eq:phat}. {In doing so, we are taking into account the fact that the $\alpha\rightarrow 0$ limit is well defined, motivating the choice $t=0$ made before expression \eqref{eq:2snyder}. In other words, we are performing a deformation of Snyder theory to curved spaces: we can recover the Snyder case in this limit, while generally we obtain some additional contributions in powers of $\alpha$. }{Actually, as explained after eq. \eqref{eq:2snyder}, one could have started with a theory in a de Sitter background, where the correct kinetic operator would be $A_{dS}=\hat{p}^2+\frac{\alpha^2}{2} J^2_{ij}$, $S_K$ would have had a nontrivial measure and then one could have thought of introducing a noncommutative deformation through the parameter $\beta$. One may also define an action invariant under the full Snyder-de Sitter group, and hence under duality, which seems to be rather involved. We shall consider some of these points in a forthcoming publication. }

Using hermitian operators $\hat{x}$, $\hat{p}$ as given by eqs.~\eqref{eq:symmetrization} and \eqref{eq:phat}, and  after several manipulations using the commutators of the algebra and integration by parts, we can obtain a simpler expression, that up to first order in both $\alpha^2$ and $\beta^2$ reads
\begin{align}\label{kinetic}
 \begin{split}
\hat p^2&\approx p^2+\frac{\alpha^2}{\beta^2}x^2
 +\alpha^2 (x_ix_jp_jp_i+x_jp_jp_i x_i)\\
 &= p^2+\frac{\alpha^2}{\beta^2}x^2
 +\alpha^2_{\text{eff}} x_jp_jp_ix_i-\frac{\alpha^2}{2} D(D+1),
 \end{split}
\end{align}
in terms of $\alpha_{\text{eff}}^2=2\alpha^2$. 
It is interesting to notice that as a by-product we get a negative contribution to the mass term, providing an effective mass 
\begin{align}
m_{\text{eff}}^2=m^2-\frac{\alpha^2}{2} D(D+1). 
\end{align}
Even if this could seem awkward at first sight, the appearance of such a term and the usually consequent infrared divergences are familiar in the commutative de Sitter case \cite{Marolf:2012kh}. It is not the aim of this paper to inquire in this direction. Instead, we point out that there have been some advances to show {that this is a pure perturbative feature}-- this can be inferred for example from the computations in the large $N$ limit of a $O(N)$ model in commutative de Sitter \cite{LopezNacir:2019ord} {or from the use of nonperturbtative RG techniques in de Sitter\footnote{There is a subtlety: $x_jp_jp_ix_i$ is Hermitian, but not necessarily positive. However, acting as $\int \phi (x_jp_jp_ix_i\phi)$, is positive. }\cite{Moreau:2018lmz}}.

Hence, up to first order order in the noncommutative parameters one obtains the Grosse-Wulkenhaar kinetic term, plus an extra contribution that mixes position and momenta operators.
It is important to notice that in this way the harmonic term, added ad hoc in the Grosse-Wulkenhaar model, acquires a {clear geometric and physical} meaning. This possibility was proposed in a different context already in \cite{Buric:2009ss}.

It is also interesting to remark that the form of the additional contribution $(x\cdot p)^2$ is not surprising, since in the usual commutative theory on curved spaces there would also be similar contributions. Furthermore we expect that when considering the dS invariant kinetic term further corrections of this type will arise.

Now that we have obtained an expression for the kinetic term, we need to introduce the interaction. In order to do this, let us briefly mention some properties of a powerful tool used to describe noncommutative theories, the star product. The noncommutative geometry can be implemented in terms of an algebra of functions with a deformed (noncommutative or star) product. In the case of the explicit realization we are using for the Snyder space, the star product $\star$ of plane waves has been found to be \cite{Meljanac:2017ikx}
\begin{align}
 e^{ik\cdot x}\star e^{i q \cdot x}=\frac{e^{iD(k, q)\cdot x}}{(1-\beta^2 k\cdot q)^{(D+1)/2}},
\end{align}
where $D$ is the dimension of the space and
\begin{align}
 D_{\mu}(k,q):=\frac{1}{1-\beta^2 k \cdot q} \left[\left(1- \frac{\beta^2 k \cdot q}{1+\sqrt{1+\beta^2 k^2}}\right)k_{\mu}+ \sqrt{1+\beta^2 k^2} q_{\mu}\right].
\end{align}
Note that this star-product is both noncommutative and nonassociative.

With the aid of this star-product the addition of an interacting term is straightforward. If we propose a quartic term, we end up with the usual expression in Snyder space, where the noncommutativity among position operators can be traded for the star product \cite{Franchino-Vinas:2018jcs}:
\begin{align}\label{eq.interaction}
\begin{split}
 S_I&=\frac{\lambda}{4!}\int d^Dx \; \varphi(\hat{x}) \big[ \varphi(\hat{x}) \big(\varphi^2(\hat{x})\big)\big]\\
 &=\frac{\lambda}{4!}\int d^Dx \; \varphi(x) \star \big[ \varphi(x) \star \big(\varphi(x) \star \varphi(x) \big)\big].
 \end{split}
\end{align}
Notice that the original duality present in the Snyder-de Sitter algebra, viz. eq. \eqref{eq:duality}, is broken by the inclusion of such a potential. In order to preserve it, one should introduce a noncanonical potential including both $\hat x$ and $\hat p$ on equal footing, what seems to us not strongly motivated.

Going back to formula \eqref{eq.interaction}, a direct computation gives an expansion for small noncommutativity $\beta$,
\begin{align}
 S_I&=\frac{\lambda}{4!} \int d^Dx \,\left[\varphi^4 + \beta^2\, \varphi_{\star,(1)}+\mathcal{O}(\beta^4)\right],
\end{align}
where we have defined the first noncommutative contribution to the potential
\begin{align}\label{eq.phi_fourth}
 \varphi_{\star,(1)}:&= \frac{2}{3} \varphi^3    \Big( (D+2) +2 x^{\mu} \partial_{\mu} \Big) \partial^2\varphi.
\end{align}
{This expansion will become relevant when considering the divergences that will arise in the following section,
where we compute the one-loop effective action. }

{As the reader could guess, in order to make the computations feasible,
the starting action whose  one-loop contributions will be considered in the next section is simply the sum of  both the kinetic and interacting term, }
\begin{align}\label{eq.S}
 S=S_K+S_I,
\end{align}
{in the regime where both the curvature $\alpha$ and noncommutativity $\beta$ parameters are small.}

\section{The one-loop effective action}
In order to perform the one-loop calculations we will use the Worldline Formalism in its noncommutative version \cite{Bonezzi:2012vr}.
This method has already proved its utility in the study of the
exact nonperturbative propagator of the Grosse-Wulkenhaar model \cite{Franchino-Vinas:2015vpa, Vinas:2014exa}.

Consider then the expression for the effective action $\Gamma$ up to one-loop corrections,
\begin{align}
\Gamma[\phi] = S[\phi]-\frac{1}{2} \int_{0}^{\infty}\frac{dT }{T}\text{Tr} \left( e^{-T \,\delta^2 S}\right) ,
\end{align}
where the connection with the heat kernel of the second variation $\delta^2 S$ of the action is made explicit. The fact that the action involves the noncomutative and nonassociative product $\star$ implies that the computation of the variation should be performed with care. As an example of this type of calculations consider \cite{Franchino-Vinas:2018jcs}.

Instead of writing the second variation of the action, we report below its Weyl ordered expression $\delta^2 S_{W}$, since this is the relevant one for the Worldline Formalism:
\begin{align}\label{eq.S_W}
 \delta^2 S_{W}=p^2+\omega^2x^2
 +\alpha_{\text{eff}}^2\, (x_ix_jp_jp_i)_S+m^2+V_W,
\end{align}
where the subscript $W$ indicates a Weyl-symmetrized expression\footnote{As an example, consider for simplicity the one-dimensional case. If we have an expression $x^2 p^n$, its symmetrization gives $\left(x^2 p^n\right)_S=\frac{1}{4}\left(x^2 p^n+2xp^nx+p^nx^2\right)$ \cite{Bastianelli:1992ct}.} and we have introduced the Weyl-ordered potential $V_W$ and  the frequency parameter $\omega$ for the oscillator\footnote{Note that $\omega$ is actually a frequency times a mass, that can be identified with the mass $m$ of the field.},
given by\footnote{At this point $\phi$ is the classical field.}
\begin{align}\label{eq.vW}
\begin{split}
 V_{W}:&=\frac{\lambda}{4!}\frac{1}{2}\int  \frac{dq_1 dq_2}{(2\pi)^{2D}} \left[4! e^{ i x ( q_1+q_2)} +\beta^2 \left(\alpha_{\mu\nu}(x) p^{\mu} p^{\nu}+\beta_{\mu}(x) p^{\mu}+\gamma(x)\right)\right] \tilde{\phi}_{1}\tilde{\phi_2},\\
 \omega^2:&=\frac{\alpha^2}{\beta^2}.
 \end{split}
\end{align}
In the expression for $V_W$, the $x$-dependent coefficients $\alpha_{\mu\nu}$, $\beta_{\mu}$ and $\gamma$ are those given in Appendix A of \cite{Franchino-Vinas:2018jcs}.
{In the following, we shall consider $\omega$ as an independent parameter, since its renormalization flow is different from that of $\alpha/\beta$}

Using expressions \eqref{eq.S_W} and \eqref{eq.vW}, the computation of the one-loop contribution to the effective action is lengthy, but otherwise straightforward. In order to isolate the divergent contributions and proceed with the renormalization of the theory, we use dimensional regularization in $D=4-\varepsilon$ dimensions. In this way we find that the divergent contributions of the 2- and 4-point functions are given by
 \begin{align}\label{eq.2point_div}
\Gamma^{(2)}_{\text{div}}&=-\frac{\lambda}{192\pi^2\varepsilon }\int d^Dx \,\phi \,\Bigg[-26 \alpha_{\text{eff}}^2 + 6 m^2 + 12 m^4 \beta^2 + \frac{6 \alpha_{\text{eff}}^2 m^4}{\omega^2} + 48 \beta^2 \omega^2 
\\
&\hspace{1cm}+ x^2 (15 \alpha_{\text{eff}}^2 m^2 + 6 \omega^2 + 36 m^2 \beta^2 \omega^2 - 4 \beta^2 \omega^2 \partial^2) - 8 \beta^2 \omega^2  x_{\mu} x_{\nu} \partial{}^{\mu} \partial^{\nu}
\\
&\hspace{8cm}+ x^4 (9 \alpha_{\text{eff}}^2 \omega^2 + 24 \beta^2 \omega^4) \Bigg]\phi,\nonumber\\
  \Gamma^{(4)}_{\text{div}}&= \frac{1}{4!}\frac{3  \lambda ^2 }{16 \pi ^2 \varepsilon}\int d^Dx\,\phi^2\, \left[ \frac{\alpha_{\text{eff}} ^2 \partial^2}{2 \omega ^2}-\frac{ 2 m^2  \left(\alpha_{\text{eff}}´ ^2+2 \beta ^2 \omega ^2\right)+\omega ^2 }{\omega ^2}-\frac{ x^2 \left(5 \alpha_{\text{eff}} ^2+16 \beta ^2 \omega ^2\right)}{2 } \right.\Bigg] \phi^2 \label{eq.4point_div}\\
  &\hspace{3cm}-\beta ^2 \phi_{\star,(1)},\nonumber
\end{align}
where the notation of eq.~\eqref{eq.phi_fourth} has been employed.
This result agrees with \cite{Franchino-Vinas:2018jcs} once we first
take the limit of $\alpha$ going to zero and then consider a vanishing $\omega$.
{Notice that higher n-point functions are not considered here, since it is not our purpose to analyze the renormalizability of the model.}

Some remarks are now necessary concerning the origin of these divergent terms.
In order to proceed with the renormalization of the theory, the need to introduce counterterms
for the effective mass, the frequency and the coupling constant should be evident.
Moreover, the appearance of a $x^4$ term in the 2-point function and a $x^2$ term in the 4-point function
can be motivated by considering the inclusion of the determinant of the metric in the original action --
indeed, in de Sitter projective coordinates this is given by \cite{Ivetic:2013yga}
\begin{align}
\sqrt{\det g}=\frac{1}{(1+\alpha^2 x^2)^{(D+1)/2}},
\end{align}
which upon expansion in the $\alpha$ parameter gives (among others) the desired contributions. So the introduction
of this type of counterterms should not be seen as an inconsistency of our theory.

Another contribution related to the geometry of the curved space are the two terms
combining $p$ and $x$. As explained in the previous section,
they are of the form that usually arises when considering the Laplacian in curved commutative spaces.
The fact that we should add a new term $x^2p^2$ to our original action can be understood as { a backreaction of the
field on the geometry}, i.e. a
dynamical deformation of the original spacetime by the one-loop contributions of the scalar field.
A similar effect could be seen in the commutative case, where the one-loop effects of the scalar field
generate new terms in the gravitational sector \cite{deBerredoPeixoto:2003hv, Gorbar:2003yp, Franchino-Vinas:2018gzr,Franchino-Vinas:2019upg}.

The remaining contributions, i.e.\  the $\phi_{\star,(1)}$ and the $\partial^2$ one in the 4-point function,
can be explained if we look deeper into the noncommutative $\phi_{\star}^4$ potential.
{In fact, by taking a closer look at eq.~\eqref{eq.phi_fourth} one realizes
that these two divergent contributions were hidden in our original action.}
As a consequence, the noncommutative parameter $\beta$ should be renormalized,
but in a different way for the two terms in the RHS of eq.~\eqref{eq.phi_fourth} --
in other words, we are in presence of a dynamical deformation of the noncommutative space.
This is one of the key differences with the Grosse-Wulkenhaar model, in which the
noncommutativity parameter has no running.

\section{The beta functions}
Upon the introduction of an energy scale $\mu$ to render the coupling $\lambda$ dimensionless in $D$ dimensions, we can compute the corresponding beta functions from \eqref{eq.2point_div} and \eqref{eq.4point_div},  using the standard definition $\beta_{x}=\frac{\partial x}{\partial \log \mu}$ for the coupling $x$. For the sake of simplicity we limit ourselves to the terms that were present in the original action \eqref{eq.S}, whose coupling constants have the following beta functions:
\begin{align}\label{eq.system}\begin{split}
 \beta_{\lambda} &= \frac{3 \lambda^2}{16 \pi^2 } \left(1 + 4 m^2 \beta^2+\frac{2 \alpha_{\text{eff}}^2 m^2}{\omega^2}  \right),
 \\
 {\beta}_{{\omega}^2}&=\frac{\lambda}{96 \pi^2} (15 \alpha_{\text{eff}}^2 m^2 + 6 \omega^2 + 36 m^2 \beta^2 \omega^2),\\
 {\beta}_{{m}^2_{\text{eff}}} &=\frac{\lambda}{96 \pi^2} \left(-26 \alpha_{\text{eff}}^2 + 6 m^2 + 12 m^4 \beta^2 +  \frac{6 \alpha_{\text{eff}}^2 m^4}{\omega^2} + 48 \beta^2 \omega^2\right),\\
 {\beta}_{{\alpha}_{\text{eff}}^2} &=\frac{\lambda }{12 \pi^2}\beta^2 \omega^2,\\
 {\beta}_{{\beta}^2}&=-\frac{3\lambda }{8 \pi^2 }  m^2 \beta^2 \left(\frac{\alpha_{\text{eff}}^2}{\omega^2} + 2 \beta^2 \right).
 \end{split}
\end{align}
Contrary to the GW case, the field does not get renormalized at the one-loop level.

\begin{figure}
\begin{center}

 \begin{minipage}{0.49\textwidth}
 \includegraphics[width=1.0\textwidth]{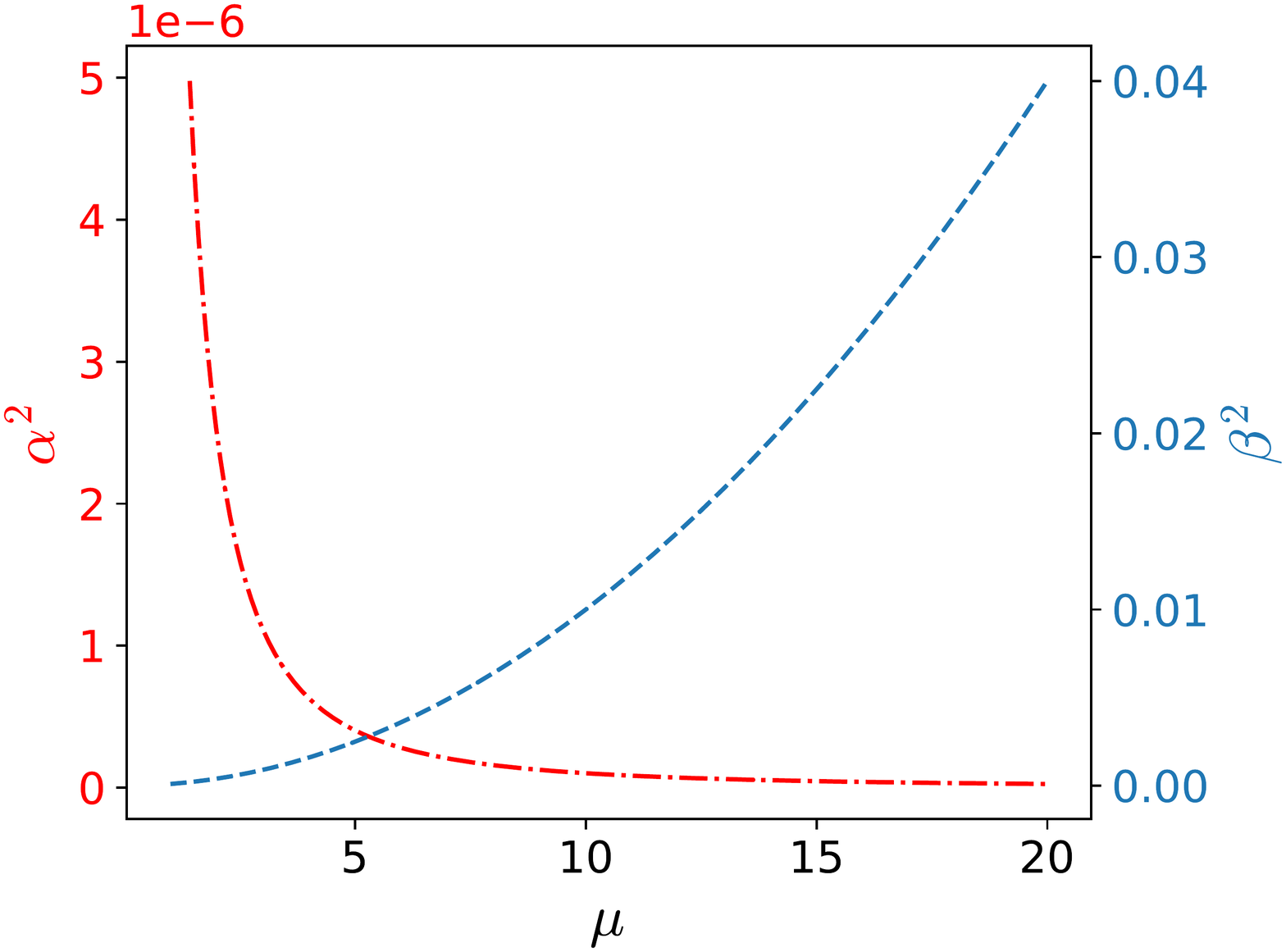}
 \end{minipage}
 \begin{minipage}{0.49\textwidth}
 \includegraphics[width=1.0\textwidth]{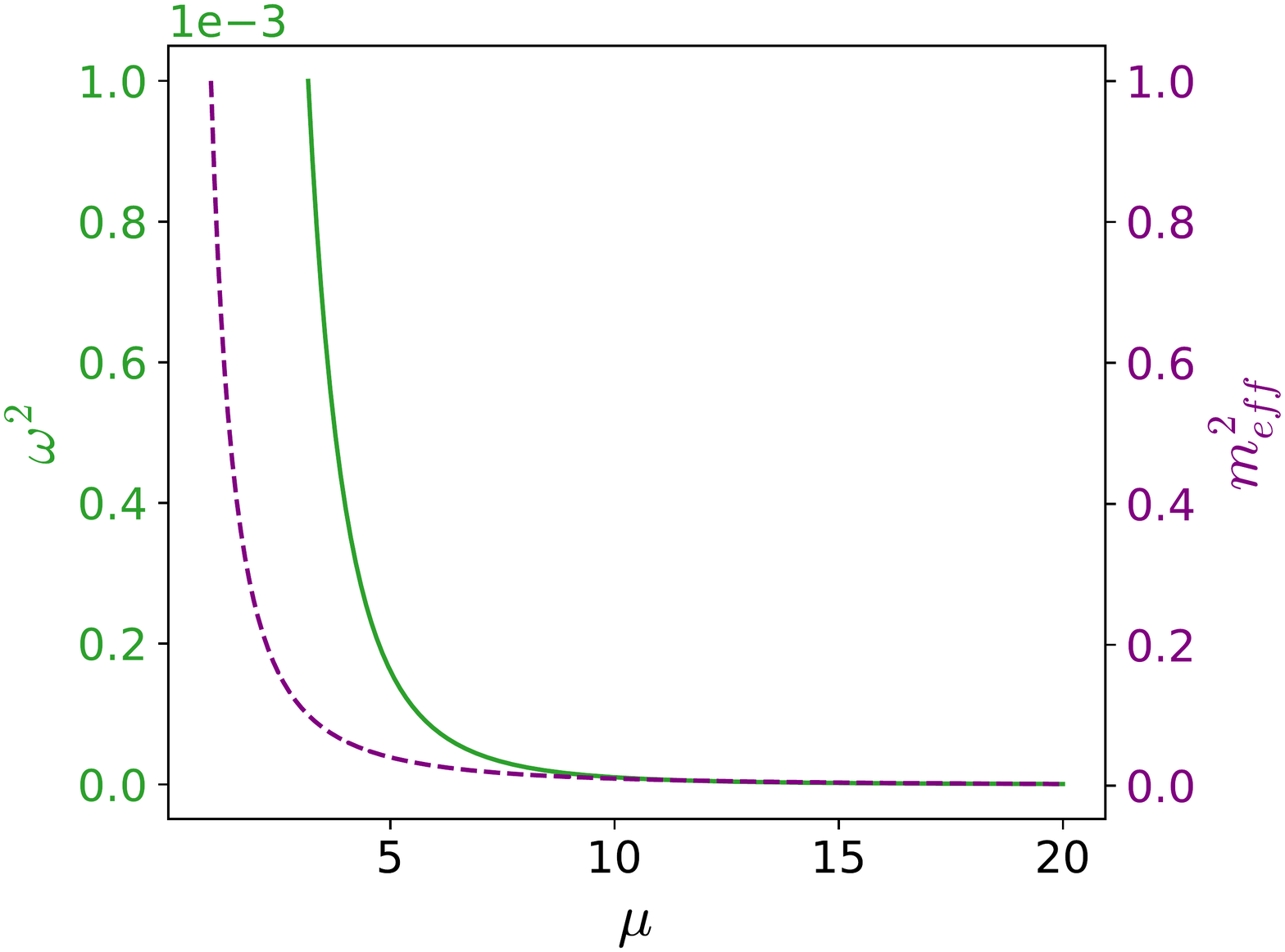}
 \end{minipage}

 \begin{minipage}{0.49\textwidth}
 \includegraphics[width=0.9\textwidth]{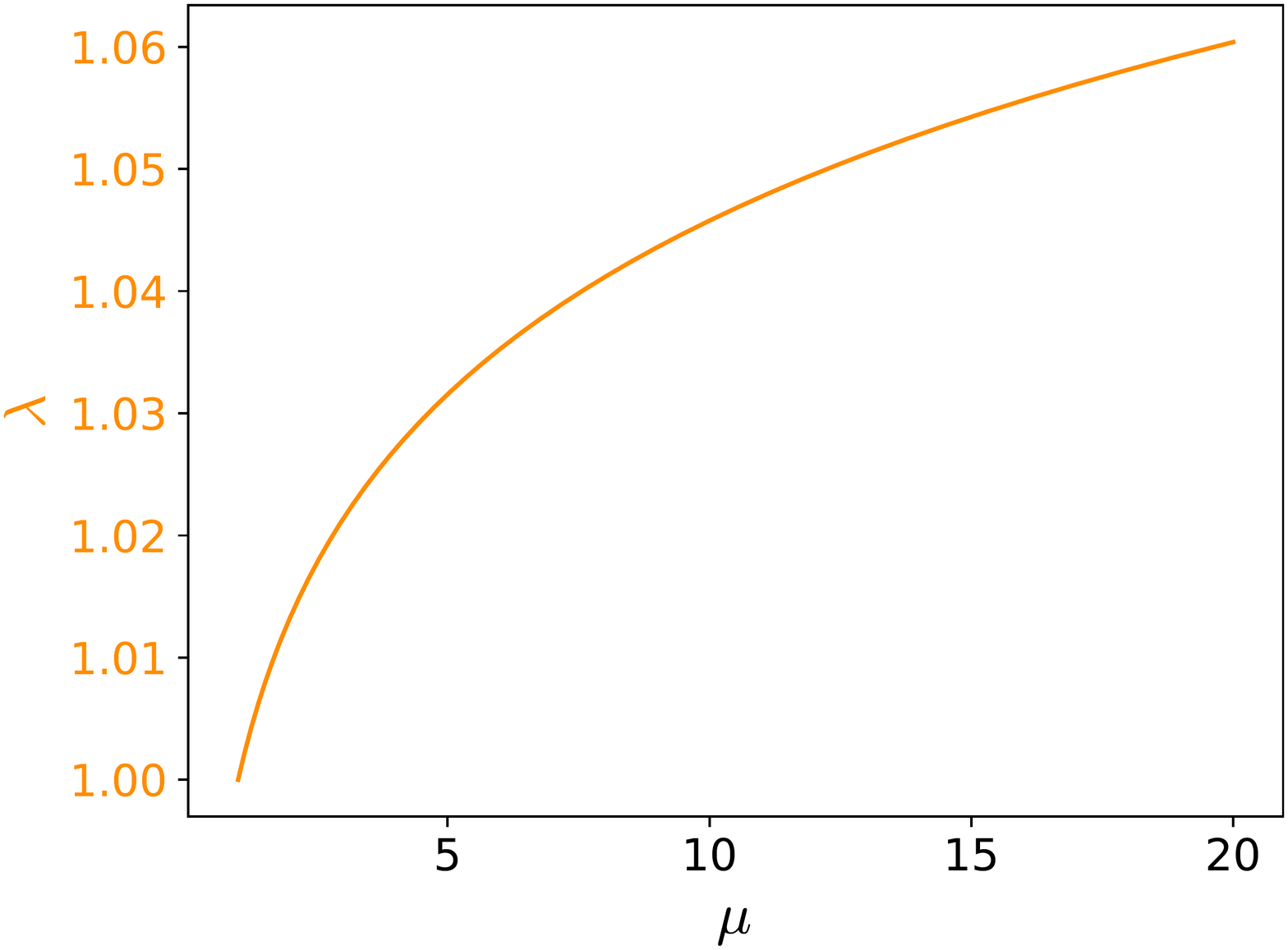}
 \end{minipage}
 \caption{Numerical solutions for the running of all the parameters of the theory, choosing as initial conditions $v_1=\left(1,\, 1,\,10^{-5},\,10^{-4},\,10^{-1}\right)$
 at an energy $\mu_0=1$ (all the quantities are measured in the corresponding powers of $\text{eV}$). }
 \label{fig.functions}
 \end{center}
\end{figure}

Recall that, in order to analyze the renormalization group equation, the relevant differential equations are obtained
from the previous ones adding a term related to the dimension $d_{x}$ of the corresponding coupling $x$ in units of mass\cite{peshkin}, i.e.
\begin{align}\label{eq.dimensions}
 \frac{\partial \bar{x}(\mu)}{\partial \log \mu}= \beta_{x}-d_{x}\bar{x}(\mu).
\end{align}
If one were to solve this system of differential equations, one would need to provide some initial conditions; the natural guess
would be to fix them at an energy scale attainable experimentally, say $\mu_0\sim \text{GeV}$,
at which according to the available experimental data we could choose $\lambda_0\sim1$, a typical baryon mass $m_{eff,0}\sim \text{GeV}$,
$\alpha_0\sim10^{-33} \text{eV}$ (according to its relation to the cosmological constant) and $\beta_0\sim 10^{-29} \text{eV}^{-1}$, that is of the order of the Planck scale.
These quantities fix the initial value of the frequency parameter $\omega_0\sim 10^{-4} \text{eV}^2$ in natural units, which turns out to be the only relevant effect
of noncommutativity and curvature until one reaches large energies.

Since we are interested in the qualitative behavior of the solutions, we will instead choose
rather general initial conditions. In addition, we will use
the vectorial notation $v=\left(\lambda,\,m_{eff},\,\alpha^2,\,\beta^2,\,\omega^2 \right)$ to simplify the following description.

Let us start by choosing $v_1=\left(1,\, 1,\,10^{-5},\,10^{-4},\,10^{-1}\right)$ (all measured in the corresponding powers of $\text{eV}$) at $\mu_0$.
Using the \emph{odeint} function implemented in \cite{scipy}, we solve numerically the renormalization group equations,
i.e. the system of differential equations \eqref{eq.system}, after the inclusion of the
corresponding dimensional term \eqref{eq.dimensions}. In this way we obtain the plots
depicted in Figure \ref{fig.functions}. Notice that in some plots the running of two parameters is shown.
In order to avoid confusion, in the online version the curve and the ordinate scale of a given parameter are drawn with the same color, while for different parameters the colors differs.
In the printed black and white version, the dashed curves correspond to $\beta^2$ and $m_{eff}^2$, the dashed-dotted ones to $\alpha^2$ and the continuous ones to $\lambda$ and $\omega^2$.

Turning to the interpretation of the plots,
we observe that the couplings behave as expected: the noncommutativity becomes more relevant at high energies,
contrary to what happens to the curvature (or, analogously, noncommutativity of the momenta),
while the mass and the frequency decay quickly.
On its side, the coupling constant increases slowly but faster than in the commutative case,
since its beta function is always greater than the commutative one. As a consequence, there occurs a Landau pole for $\lambda$.

\begin{figure}
\begin{center}
\begin{minipage}{0.49\textwidth}
 \includegraphics[width=1.0\textwidth]{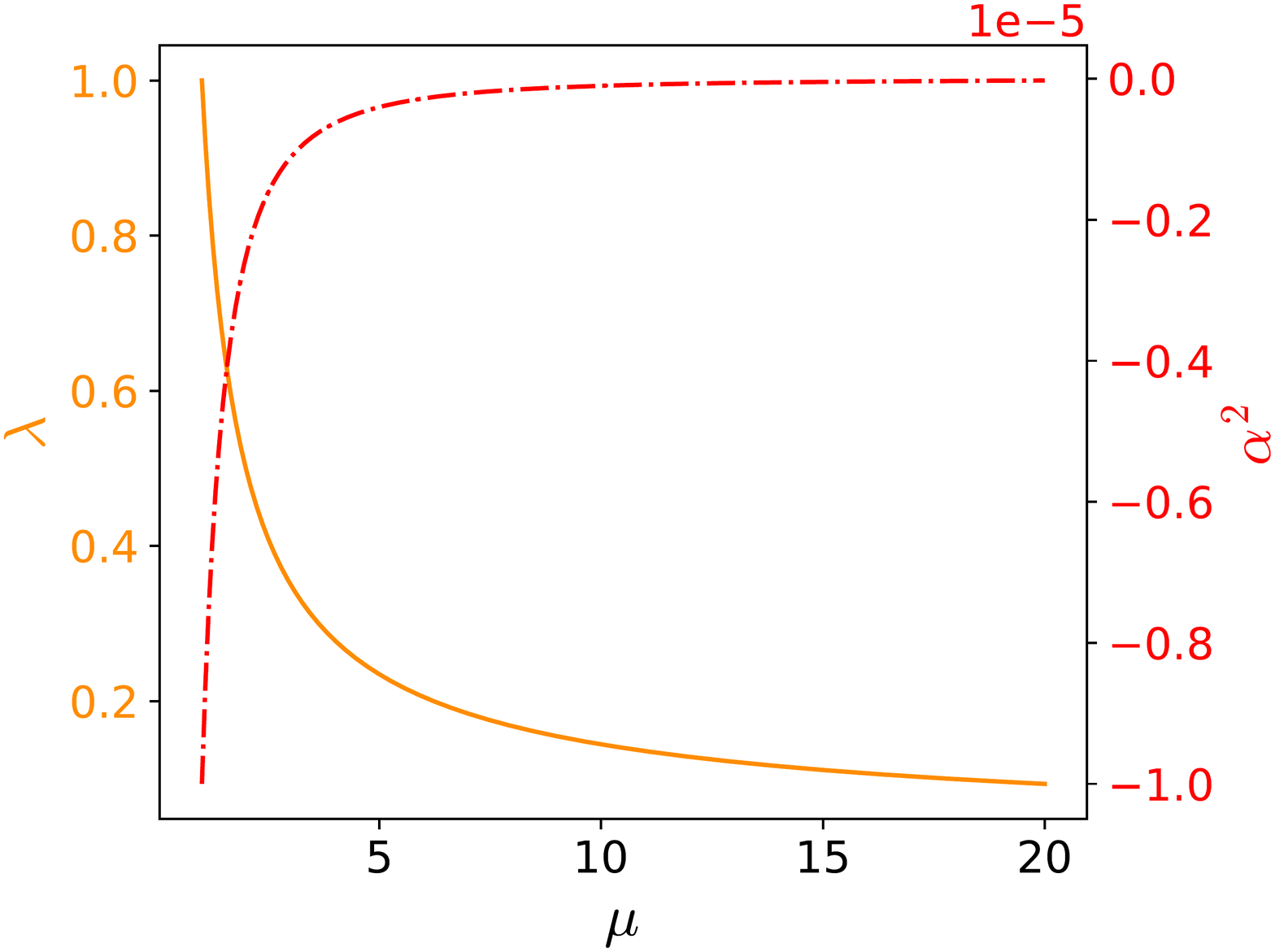}
 \end{minipage}
 \caption{Numerical solutions for the running of the $\lambda$ (continuous and orange curve) and $\alpha^2$ (dotted-dashed and red curve) parameter. The initial conditions correspond to
  $v_2=\left(1,\,10,\,-10^{-5},\,- 10^{-1},\, 10^{-1}\right)$ at the initial energy $\mu_0=1$
(all the quantities are measured in the corresponding powers of $\text{eV}$).}
 \label{fig.functions4}
\end{center}
\end{figure}

However, the situation can change drastically if we consider different initial conditions.
Suppose that we had begun with an anti-Snyder ($\beta^2<0$) and anti-de Sitter ($\alpha^2<0$) space,
given by $v_2=\left(1,\,10,\,-10^{-5},\,- 10^{-1},\, 10^{-1}\right)$ and depicted in Figure \ref{fig.functions4}.
Being the curvature and noncommutative contribution to $\lambda$'s beta function negative,
the asymptotic freedom of the theory is guaranteed for large enough masses. Also, as can be seen from Figure \ref{fig.functions4},
the curvature tends to zero for large energies as in the previous case -- analogously, the frequency, the mass and the noncommutative parameter show a behaviour analogous to the previous case
(of course with an additional minus sign for $\beta^2$). All in all, in this situation the only relevant parameter in the UV turns out to be the noncommutativity parameter.

\begin{figure}
\begin{center}
\begin{minipage}{0.49\textwidth}
 \includegraphics[width=1.0\textwidth]{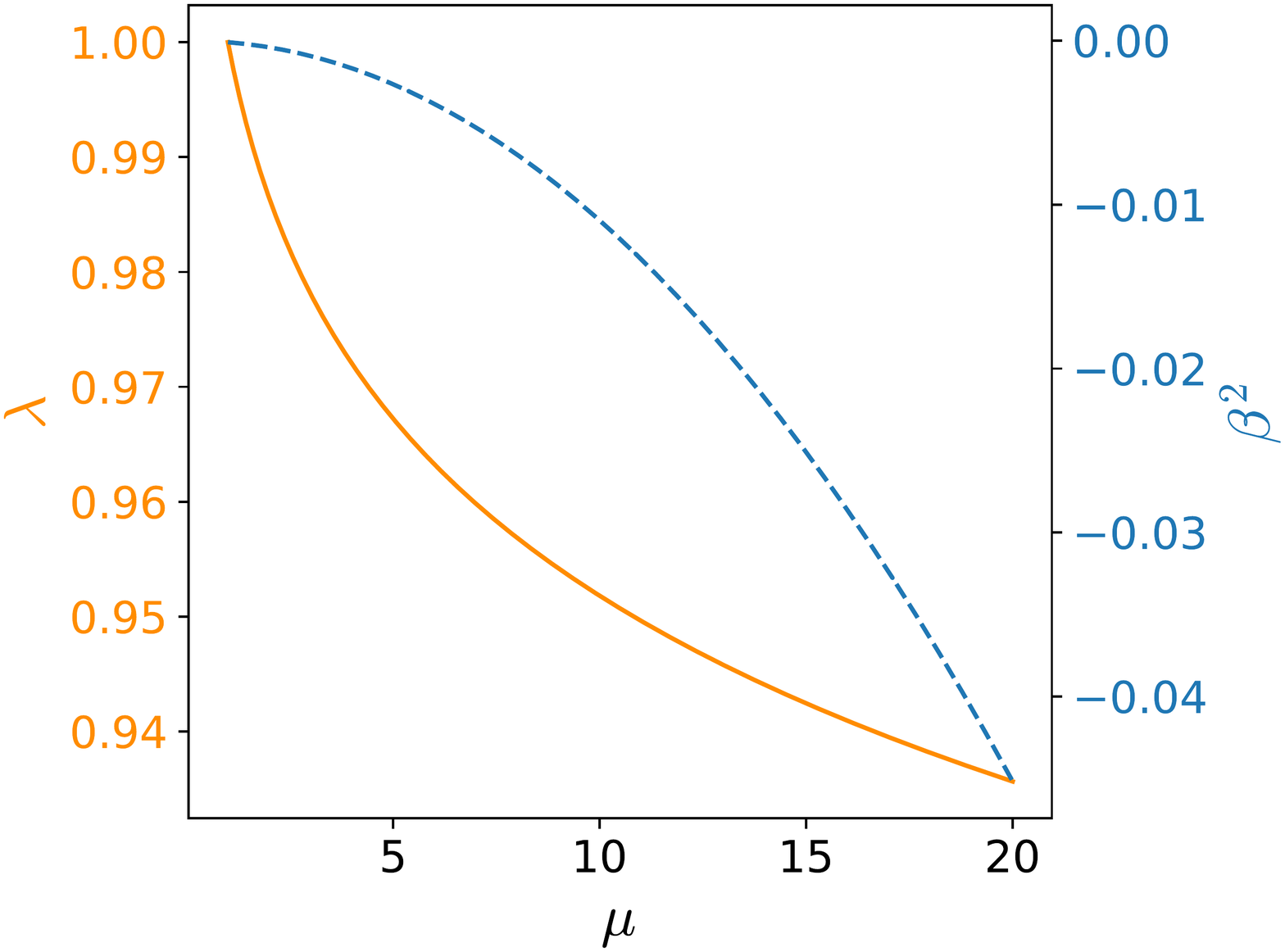}
 \end{minipage}
 \begin{minipage}{0.49\textwidth}
 \includegraphics[width=0.9\textwidth]{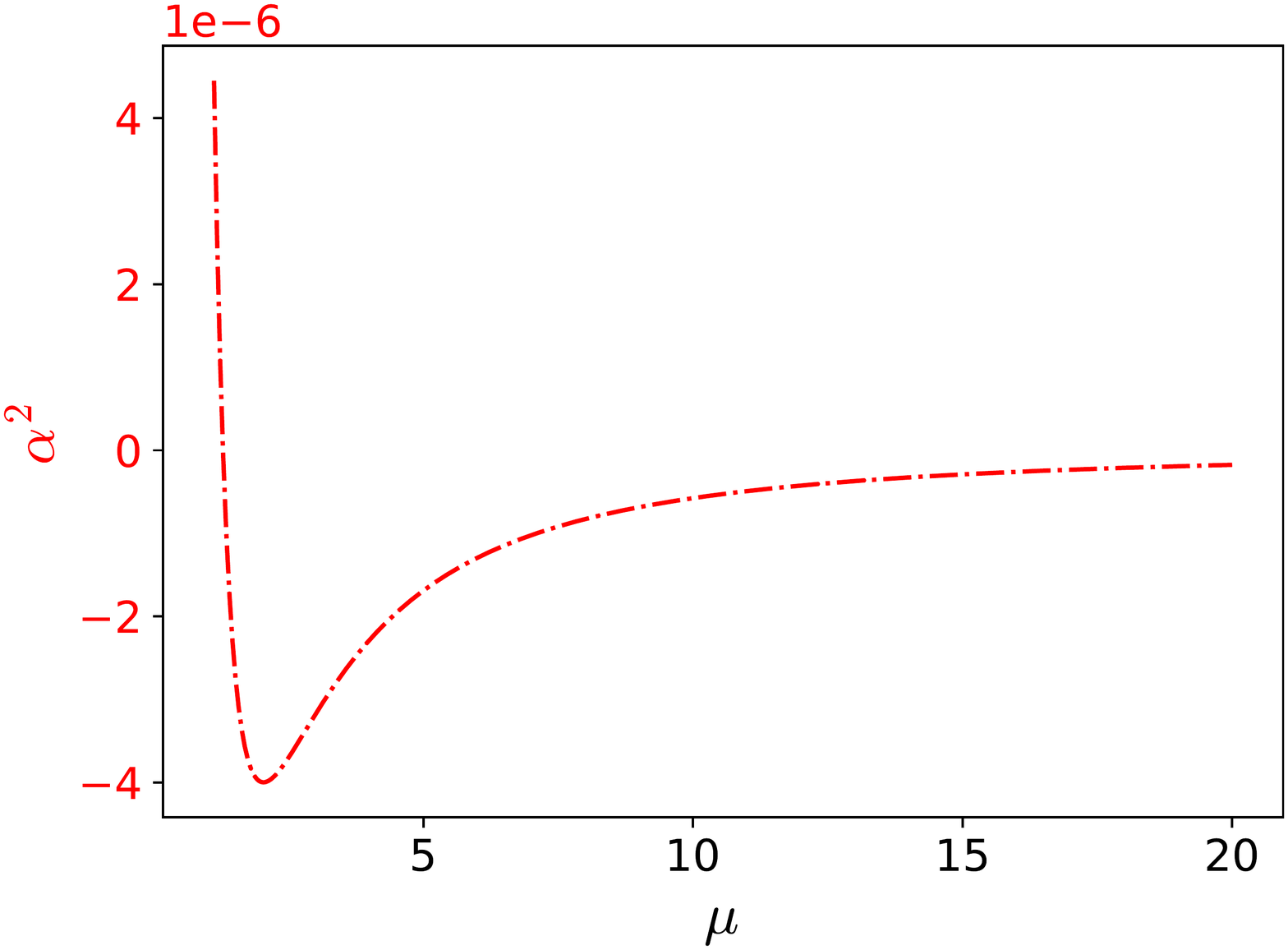}
 \end{minipage}
 \caption{Numerical solutions for the running of $\lambda$ (continous and orange curve), $\beta^2$ (dashed and blue curve) and $\alpha^2$ (dashed-dotted and red curve). The initial conditions correspond to
 $v_3= \left(1,\,10^2,\,10^{-5},\,- 10^{-4},\, 10^{-1}\right)$ (left panel) and $v_4=\left(1,\,10,\,10^{-5},\,- 10^{-1},\, 10^{-1}\right)$ (right panel), choosing the initial energy $\mu_0=1$
(all the quantities are measured in the corresponding powers of $\text{eV}$).}
 \label{fig.functions3}
\end{center}
\end{figure}

Now notice that, after the running, the constraint for $\omega$ given in \eqref{eq.vW} is no longer valid. This suggests that we could also
relax the constraint in the initial condition
and consider more general ones. For example, let us choose an initial anti-Snyder space,
say $m_{eff}=1$ and $\beta^2=-10^{-4}$ in the corresponding units.
Then, $\lambda$'s beta function would have a negative contribution that
would be nevertheless supressed by the fast decay of the mass. If we choose instead the situation with a bigger mass,
e.g. choosing $v_3=\left(1,\, 10^2,\,10^{-5},\,-10^{-4},\,10^{-1}\right)$,
the negative term in the beta function of $\lambda$ is dominant and the theory becomes again asymptotically free. The plots in Figure \ref{fig.functions3}
(left) depict the behaviour of $\lambda$ and $\beta^2$ in this case, while the remaining parameters show  a behavior qualitatively equal to the previous $v_1$ case.

Still another interesting situation shows up if we still assign initial anti-Snyder conditions,  but with
$v_4= \left(1,\,10,\,10^{-5},\,- 10^{-1},\, 10^{-1}\right)$ at $\mu_0$ (all quantities given in the corresponding powers of eV).
In this case $\alpha^2$ decreases so fast that becomes negative -- our interpretation is thus that
the geometry of the model becomes of anti-de Sitter type as a consequence of the one-loop dynamics of the theory. At the same time, the coupling constant
stays in an asymptotic-free regime. The only parameter that changes its behavior with respect to the previous case is $\alpha^2$, whose plot is shown in the right panel of Figure \ref{fig.functions3}.

\section{Conclusions}
The formulation of a scalar field theory on noncommutative curved spacetime, {in the approximation of small
noncommutativity and curvature, has revealed several }interesting features.
In particular, we have shown that the Snyder-de Sitter model could provide an {astrophysical} motivation for the
harmonic term introduced by Grosse and Wulkenhaar in their celebrated model \cite{Grosse:2003nw}.
In this respect, the emergence of a frequency given by the quotient of the noncommutativity and curvature parameter
turns out to be even more interesting, since it means that its effects could be several orders of magnitude bigger
than expected. Of course it will be of interest to analyze how the current experimental data from the standard model
could constrain the presence of such an harmonic term.

Turning our attention to the beta functions, we have seen that according to the initial infrared initial conditions,
it could be the case that the curvature and the noncommutativity render the theory asymptotically safe. For this effect to arise,
it is a necessary condition to begin with an anti-Snyder-anti-de Sitter or an anti-Snyder-de Sitter scenario.
In these cases it is clear that the theory does not suffer the illness of a Landau pole. However, it is important to
emphasize  that this is a situation different from
the one set by the GW model, in which the existence of a zero of a beta function guarantees asymptotic safety.

Another interesting peculiarity is the fact that we are forced to consider the running of both the noncommutativity and curvature
parameters.
Its consequence is that an initial symmetry, encoded in the constraint $\omega^2=\alpha^2/\beta^2$, is dynamically broken. 
{This in turn generates the question whether there could exist some (hidden) symmetry that could prevent the independent running of $\omega$, 
for example after the consideration of the whole expansion in $\alpha$ and $\beta$ or of a fully invariant action. Also the inclusion of a noncommutative
action for the  gravity sector might be relevant. Indeed, such a symmetry could be the equivalent of gauge symmetry in QED, 
which through the Ward-Takahashi identity precludes a different running for each of the couplings identified with the electric charge at tree level.}

{Aditionally, noncommutativity on curved spaces} could give the opportunity to analyze some proposals discussed in the literature of quantum gravity and asymptotic
safety, such as \cite{deAlwis:2019aud}. Indeed, we have shown that a running from de Sitter space to a would-be anti-de Sitter one
is allowed in our model after generalizing the initial conditions, suggesting that, even if the
de Sitter-Swampland conjecture were true, a way out of it could be possible.

Finally, we would also like to {mention some open questions. The discussion of whether
the Snyder-de Sitter model in the small $\alpha$ and $\beta$ regime is UV complete or
(all-order) renormalizable, would involve a
deeper analysis of higher n-point functions and will be thus left to future work. To this end, we
believe it { could be essential to have an equivalent of the Langman-Szabo duality present in the GW case, which could not only simplify the 
computations, but more importantly prevent some radiative corrections and avoid the need to include an infinite number of new terms in the action.\\
Moreover,} an analysis of the UV-IR divergences seems to be impossible in our framework.
This would in fact require the investigation of the theory beyond linear order in $\alpha$ and $\beta$, as initiated in \cite{Meljanac:2017jyk}.
Another interesting point regards gravitational corrections to the beta functions of coupling constants.
In some papers (see for example \cite{Gonzalez-Martin:2017bvw} and references therein), it has been proposed that this type
of corrections is not physical by analyzing QFT in curved spaces.} Our model could be a playground for considering this claim
in a theory that can be seen as a step forward towards Quantum Gravity.

\medskip
\noindent\textbf{Acknowledgements}:
SAF acknowledges support from ACRI and INFN under the Young Investigator Training Program 2018,
Functional and Renormalization-Group Methods in Quantum and Statistical Physics, and Project 11/X748 from the UNLP.
SAF would like to thank the Università di Cagliari for its hospitality.
The authors thank H. Grosse and  F. Murgia, and especially L. Zambelli for his useful help at many stages
of this work.
{ The authors would like to acknowledge networking support by the COST Action CA18108.}

\appendix

\printbibliography

\end{document}